# A literature review: What exactly should we preserve? How scholars address this question and where is the gap

Jyue Tyan Low

Phone: 1.412.692.0178
Email: folojoto@gmail.com

This paper was first written for the Spring 2011 digital preservation class of the Master of Library and Information Science program at the University of Pittsburgh, Pennsylvania, USA. This is published in the author's own capacity.

## 1.    Introduction

In discussing what attributes of digital materials to preserve, Ross Harvey, in his book, *Preserving Digital Materials* observes:

> As well as gaining a deeper understanding of the changes that we make, we need to know more about exactly what it is we are trying to preserve. What characteristics, attributes, essential elements, significant properties – many terms are used – of digital materials do we seek to retain access to? What is the 'essence' of digital materials? Whereas this question was a simple one to answer in the non-digital context, where typically we sought to conserve and preserve the original artifact, for digital materials it is not so straightforward. (76)

Digital materials are impalpable and unperceivable. They only become perceivable by humans when they manifest themselves as intelligible objects e.g., as a "document" on the computer screen, or a printout on paper. However, when discussing digital preservation, it's already an axiom that paper is a poor surrogate of the electronic document we see on screen, which can be manipulated. If printing on paper is a non-solution, how do we ensure that the "document" can always manifest itself on our screen in face of our fast-changing digital world where we are not only confronted with hardware obsolescence but inexorable software changes? And what exactly do we have to preserve to ensure that the "same" electronic document will always be evoked when we need it at some point in future?

There are generally two approaches to long-term preservation of digital materials (Borghoff et al. 2006, 13): one involves preserving the object in its original form as much as possible along with the accompanying systems, another [generally known as migration (or transformation) (Giaretta 2009)] involves transforming the object to make it compatible with more current systems but retaining its original "look and feel." Today, migration remains the most widely applied and promising approach despite the inevitable changes to the original digital object it involves (Borghoff et al. 2006; Ferreira et al. 2007; Wilson 2007). The community has come to accept the inevitable changes even though it seems to run antithetical to the basic tenet of preservation. The next pertinent question is: if a digital object is to be migrated and thus loses some of its original properties, what essential attributes must the object bring along to its next life to still manifest like the original without losing its integrity? As Harvey puts it (2005, 93), "Not all of the elements that make up a digital object are equally important in recreating the conceptual object." So how do we decide which of these elements to keep? A case in point is the periodic table offered by Thibodeau (2002). He states that structure for example, is one essential element to retain in order to preserve the integrity of the periodic table. The font type may not be important in this case.

This review addresses the question of *what exactly should we preserve*, and how the digital preservation community and scholars address this question. How the literature was selected will first be described very briefly. I then introduce the much-abused-term "significant properties," before revealing how some scholars are of the opinion that characteristics of digital objects to be preserved (i.e., significant properties) can be identified and should be expressed formally, while others are not of that opinion. The digital preservation community's attempt to expound on the general characteristics of digital objects and significant properties will then be discussed. Finally, the review shows that while there may be ways to identify the technical makeup or general characteristics of a digital object, there is currently no formal and objective methodology to help stakeholders identify and decide what the significant properties of the objects are. This review thus helps open questions and generates a formative recommendation based on expert opinion[1] that expressing an object's *functions* in an explicit and formal way (using didactic guides from the archives community) could be the solution to help stakeholders decide what characteristics/ elements exactly we should preserve.

## 2.   Methodology

To begin, I first formulated a query statement that reads: I am interested to know if there is any (criteria/ solution/guideline/ approach) that can help one define/make decision on what (aspects/elements/characteristics/features/attributes) of a digital object to preserve.

I did not know what exactly was available out there. Therefore, I embarked on what I would deem a "task-based information exploration." I carried out searches on Pittcat, SCOPUS, ACM, and IEEE databases, using query terms and phrases such as "digital preservation," "attributes," "characteristics," "elements," and "what aspects." I scanned through the abstract of close to 250 seemingly relevant articles, and narrowed down on the most relevant ones. As I read the articles, my knowledge about the topic mushroomed, and I found the specific area of studies I was looking for—"significant properties."

Using significant properties as the departure point, I then refined my searches on the databases once again, interpreted the results and selected the most targeted articles. Websites of prominent projects on significant properties mentioned in these articles were also explored. Finally, relevant articles from bibliographies of these core articles were added.

---

[1] This opinion belongs primarily to Dr Martin Doerr, a research director in the Information Systems Laboratory and the head of the Centre for Cultural Informatics of the Institute of Computer Science, FORTH.

The domain, as I discovered, is extremely vibrant, and constantly evolving. The debates are hot and very much ongoing. However, as most of the articles and projects highlighted in this review are relatively recent, I believe I have captured the most current trends and gaps in the domain.

# 3. What exactly should we preserve? Enters "significant properties"

In order to know *what* of a digital object to preserve, special interest groups, scholars and practitioners have explored on what is called "significant properties," "essential characteristics," or "essence" of digital objects.

As mentioned in the introduction, Harvey (2005) observes that many terms have been used interchangeably to describe the characteristics/ elements of digital objects that must be retained for long-term preservation. The three formalized terms are "significant properties" used primarily by the InSPECT[2] project (Knight 2009), PLANETS[3] (Brown 2008; Becker et al. 2009), and others; "essential characteristics" used by NARA[4] until October 2009 where they then proposed the use of the term "significant properties" for consistency (NARA 2009); and "essence" used almost only exclusively by the National Archives of Australia (NAA) (Wilson 2007).

I am definitely a second messenger here, as many scholars, including Wilson (2007), as well as Knight and Pennock (2008), have chronicled the evolution of the term "significant properties," and reported on project teams that have researched and contributed over the past decade or so to the concept of significant properties.

The Cedars Project, a project which ran from April 1998 to March 2002, first coined the term "significant properties" (CEDARS 2002). However, some scholars recognize works of earlier contributors who first introduced this concept that lead to what is now known as significant properties. These scholars include Wilson (2007) who recognizes Clifford Lynch as the first to introduce the idea of canonicalization in his 1999 article[5]; and Knight and Pennock (2008), who identify Jeff Rothenberg and Tora Bikson as early scholars who first considered the use of significant properties as a digital preservation strategy[6] also in 1999.

## 3.1 Definition:

According to Giaretta (2009), there are many definitions of significant properties by many parties within the short period of ten years. For the purpose of this review, we are using—according to Dappert and Farquhar (2009), and NARA (2009)—the most widely accepted definition by InSPECT, found in Knight (2009, sec. 1.3):

> The characteristics of digital objects that must be preserved over time in order to ensure the continued accessibility, usability, and meaning of the objects, and their capacity to be accepted as evidence of what they purport to record.

---

[2] InSPECT is acronym for Investigating the Significant Properties of Electronic Content over Time. http://www.significantproperties.org.uk/
[3] PLANETS is acronym for Preservation and Long-term Access through Networked Services. http://www.planets-project.eu/
[4] The U.S. National Archives and Records Administration. http://www.archives.gov/
[5] "Canonicalization: A Fundamental Tool to Facilitate Preservation and Management of Digital Information"
[6] In their article "Carrying Authentic, Understandable and Usable Digital Records through Time: Report to the Dutch National Archives and Ministry of the Interior"

Still in the same year, Dappert and Farquhar (2009) articulated the difference between "significant properties" and "significant characteristics" and how they should not be used interchangeably. They concluded that one can only *preserve significant characteristics* but not *preserve significant properties.*[7] The purpose of their paper is thus to illuminate confusing terms (including significant properties and digital objects), which the authors said is expected in the precipitation of any new knowledge when terminologies are still being refined. After all, the field of digital preservation has only gained heightened interest in the last fifteen years, and the notion of significant properties was only conceived in the early 2000s.

On the basis of this definition, we next look at how scholars and researchers debate over whether significant properties can actually be expressed formally.

## 4. Can significant properties of digital objects be defined?

Many in the digital preservation community hold the common view (though with slight nuances) that in order to know what exactly we should preserve, one needs to first define the digital object's significant properties. Though there are still others who argue that significant properties cannot be formally determined. So here is an assimilated commentary to the two sides of the story.

Hedstrom and Lee (2002) believe we can and should express significant properties formally. They devoted their whole paper to elaborate on the purposes and practical applications of expressing significant properties of digital objects. Among all the reasons on why there should be a "formal model of significant properties" is of course the ability to identify what properties merit preservation. They also further state how knowing the significant properties of objects can serve to evaluate a preservation action, since it allows one to measure and compare any losses and their likeness to the originals. Other proponents who believe that significant properties should be defined include del Pozo, Stawowczyk, and Pearson (2010).

Deken (2004) suggests that one should define an object's significant properties because it is necessary to know *what* we are preserving (i.e., know the significant characteristics of digital objects first), before deciding *how* to preserve them. Deken (2004) provides an interesting perspective on how we cannot preserve everything and therefore need to know what the significant properties are. She illustrates how in the world of analog artifacts, a book for example, is the *product* of a *process* (i.e., printing) that involves many *parts* (e.g., author, typewriter, printing press machine). Usually, only the product (i.e., the book) is preserved as it is the main embodiment of the cultural content. The other parts may or may not be preserved (in any case, we will not be able to preserve the author!) In the digital world however, the product (e.g., the final rendition on a computer screen unless it's printed on paper) is often too tightly mashed up in time and space with that of the parts (e.g., collection of bits of the software program, collection of bits of the data, the hardware) and the process (i.e., reading of the bits). There is no way for us to isolate the product (i.e., rendition on the screen) from its parts or the process. Deken (2004, 237) posits that in order to preserve the product in a digital realm, one needs to preserve the parts and process "on an ongoing basis." This however is unrealistic and never going to be achievable in the author's opinion. Her recommendation is thus to capture only some significant properties of the "digital performance" or digital manifestation for realistic and sustainable preservation.

---

[7] According to the Dappert and Farquhar (2009), a characteristic is a property-value pair. One can thus only preserve a characteristic and not the property alone. See article for detail.

Probably the most convincing is Thibodeau (2002), who reminds us that one important goal of digital preservation, is to ensure that the information transferred over time is authentic. Thibodeau (2002) opined that the notion of digital object preservation is not so much to "retrieve" the object but to "reproduce" the same manifestation. In order to re-instantiate a performance/ output of a digital object, he hints that one must first define its significant properties in order to determine which of these elements to preserve that will evoke the same performance/ output (which then mean accessibility has been achieved). Thibodeau (2002) follows to say that a digital object can therefore be allowed to change as long as the final output is what is expected to be. The most notable proponent of this idea is the National Archives of Australia (NAA) which introduces the "performance model" and the notion of "preservable essence" in 2002 (Wilson 2007). The NAA performance model dictates that so long as the output/ performance can be achieved, not every element of a digital object needs to be preserved in their original state to warrant it authentic. Many in the digital preservation community including The National Archives (TNA) of the United Kingdom share that same notion that has just been described. TNA states that (2002, 8), "a record can be considered to be essentially complete and uncorrupted if the message that it is meant to communicate in order to achieve its purpose is unaltered." This is underpinned by a seminal piece by Rothenberg (2000, 57), who argues that there is no digital preservation strategy that will retain all aspects of a digital entity; and authenticity can be defined by "whether an informational entity is suitable for some purpose." On this basis, we can conclude that these scholars infer that one could conceivably select and need only to preserve the significant characteristics of a digital object that will deliver the performance for its intended use. The above affirms that in order to know *what exactly we should preserve* (that will preserve an object's authenticity), we need to first express significant properties formally.

The Authenticity Task Force of InterPARES (International Research on Permanent Authentic Records in Electronic Systems) also acknowledges the need to identify "elements of an electronic record" essential for verification of its authenticity (MacNeil 2000, 54). The task force has also invested in case studies to find out from creators of digital records "which specific elements [they] consider essential for verifying the record's authenticity" (MacNeil 2000, 56). Thus, as seen here, defining significant characteristics of digital records is considered as a form of conceptual requirement for the verification of authenticity of digital objects.

While many have spoken on the need to define significant properties, some remain skeptical. Bradley (2007) wrote of how "significant properties" is no longer a topic worthy of discourse. His reason is because, for one thing, the process of defining significant properties still cannot be automated. In addition, Bradley is of the opinion that talk on significant properties is analogous to attempting to address a philosophical conundrum, and cannot lead to any objective solution on what to preserve. People either end up with sweeping properties about the classes of materials that are not pointed enough for meaningful selection; or end up in a position that dictates preserving all the properties of the object.

Yeo (2010) also highlights the problems in trying to detail significant properties of digital objects. He brought up on how there are confusion in the community on whether significant properties are attributes inherent to the object or attributes given to the object based on its uses. He also speaks of how some initiatives attempt to define and automate the significant properties of a class of format (e.g., JPEG, TIFF, PNG) when the interest of user community tends to be on a particular genre of materials (e.g., images, reports, emails, thesis).

Nonetheless, it seems that the concept of significant properties is here to stay. As Wilson puts it (2007, 3), "the fundamental challenge of digital preservation is to preserve the accessibility and authenticity of digital objects . . ." I've informed at the beginning how the most commonly accepted preservation strategy, migration, does not attempt to keep all aspects of the original digital object . Defining the significant properties of objects is thus important so that one knows which of these to preserve to ensure accessibility, and which of these to preserve to ensure that authenticity of the object is still maintained in spite of inevitable changes necessary for preservation sake.

# 5.    Defining significant properties of digital objects: Projects and Initiatives

Ferreira et al. (2007) and Bradley (2007), among others, have acknowledged that articulating significant properties is a complex business not least because of the diversity of digital objects. Thus, while there are many attempts and initiatives that exist, few classes of objects have comprehensive and well-defined significant properties (Ferreira et al. 2007). Knight and Pennock (2008) also acknowledge the difficulty of clarifying significant properties. They gave a specific example, saying that "we have yet to reach the stage where a researcher … is able to define the significant properties of their digital research without ambiguity" (Knight and Pennock 2008, 172).

Despite the skepticism and purported difficulties, researchers and scholars did not stop attempting to clarify significant properties of digital objects. Here are some major initiatives. This list ranges from broad conceptual ideas on characteristics of digital objects to projects that have actually developed characterization tools and software to determine significant properties:

- One of the early pioneers is Thibodeau (2002) who construes that every digital object is composed of a physical object, a logical object, and a conceptual object. He explains that to preserve a digital object, it is necessary to know how these objects interact e.g., to restore a website, one must have all the files and sub-files and all the logical components.
- UNESCO (2003) builds on Thibodeau's (2002) idea stating that digital objects are composed of physical objects, logical objects, conceptual objects, and essential elements (UNESCO 2003, 35). UNESCO also provides some good guidelines on how to select essential elements to preserve (UNESCO 2003, 74).
- According to del Pozo, Stawowczyk and Pearson (2010, 294), digital objects have four different forms: physical medium, bitstream, intellectual content, and visual experience. In deciding what to preserve, the authors urge stakeholders to consider significant properties in each of these forms.
- JISC (Joint Information Systems Committee) funded a project to investigate the significant properties of four object types: vector images, moving images, software, and e-learning objects. The findings were reported at the *What to Preserve?: Significant Properties of Digital Objects* workshop organized by JISC, the British Library and Digital Preservation Coalition on April 7, 2008 in London (Knight and Hockx-Yu 2008).
- The U.S. National Archives and Records Administration's (NARA) Electronic Records Archive (ERA) developed a template of characteristics for different digital record types. As of October 2009, they have completed the template for email, textual records, digital photographs, databases, web pages, and GIS records (NARA 2009). These templates are supposed to apply across relatively homogeneous record type. Under each attribute types are corresponding questions that will elicit response from archivists to help them determine the significant properties i.e., discriminate which characteristics are essential and which are non-consequential for preservation (NARA 2009; see page 18 and 19).
- JHOVE[8] is a characterization tool that can automatically typify the format of common digital objects and expand on the characteristics of the object e.g., JHOVE can determine that the format of a digital object is TIFF, and list down its properties (Becker et al. 2009; Becker, Kulovits, et al. 2008; JHOVE 2009).
- The "objective tree" approach by PLANETS Plato[9] aims to capture stakeholders' requirements in a comprehensive way. It attempts to define the significant properties of objects starting with high level stakeholders' requirements (e.g., image needs to be preserved as they are) which are then cascaded down to quantifiable terms [e.g., "the image width, measured in pixel needs to remain unchanged" (Becker et al. 2009, 142)].

---

[8] JHOVE acronym for JSTOR/Harvard Object Validation Environment. http://hul.harvard.edu/jhove/
[9] A software tool that is mean to automate the preservation planning process (Becker et al. 2009).

- Also under the PLANETS project is another development known as the eXtensible Characterization Languages (XCL) that allows automatic extraction of objects' properties (Becker, Rauber, et al. 2008). The XCL is a diagnostic tool that measures the success of file format migration from one format to another. It is based on the principle that digital objects have attributes that can be expressed in an XML language termed XCDL (XCDL is one of the XCL language) (Planets-XCL project n.d.). This attribute description of one file format can then be used to check against that of another to ensure that all the characteristics of a file have been converted successfully.

- In the InSPECT project, an "object analysis" which details the "technical composition" of the object is first performed. Then a "stakeholder analysis" where stakeholders' functional requirements are identified is then carried out. These two analyses are then matched to produce a complete list of significant properties of the object. The InSPECT project has produced detailed testing reports on four digital object types: Raster images, digital audio recording, presentational mark-up, and electronic mail (Knight 2009).

# 6.  Problem: Lack of formal objective methodology to identify which characteristics are significant

After my research for this review, it dawned on me that many of the projects mentioned in section 5 might only have been successful in defining the objects' inherent attributes/ characteristics, but not necessarily their significant properties! Why? The work by Dappert and Farquhar (2009) says it all—"significance [of an object] is in the eyes of the stakeholder." By stakeholders, they mean creator, custodian, and user of the object. It is this significance that is defined by the stakeholder that would then determine what the essential properties are that needs to be preserved. Even if one manages to successfully define the characteristics of an object, the question is: How then does one decide which of these or how much of these characteristics are significant for preservation? Someone, somehow still needs to determine that.

While there are tools to define and express the (technical) properties of objects e.g., JHOVE, and even tools (e.g., PLANETS Plato) that will make recommendations on the target preservation format that would best preserve a set of desired features (e.g., if a TIFF should be converted to JPEG2000, PNG or leave it as TIFF), there is no tool that could help decision makers objectively determine what constitute this "set of desired features" and thus what losses can be tolerated (Becker et al. 2009). Hedstrom and Lee (2002) actually acknowledge and declare in their work that there is currently no methods to decide on which of these properties are significant and thus worthy of preservation. But of course, that was in 2002. Considerable progress has been made today (Becker et al. 2009; Knight and Pennock 2009).

More recently, del Pozo et al. (2010) emphasize on the need to articulate the preservation and institutional intent and how that will help in determining what to preserve. However, like most other preceding studies (e.g., Deken 2004), it remains abstract and the authors did not elaborate on how such intentions can be translated into tangible terms of what to and what not to preserve.

In this regard, the most recent and successful by far is the InSPECT project. Knight (2009) in the InSPECT framework report proposes a comprehensive way to define significant properties of objects (i.e., appraise on what to preserve) by considering both the stakeholders' requirements as well as the object's "technical composition and the purpose for which it may be used" (Knight 2009, sec. 3.1).

Another interesting approach is the PLANETS Plato's preservation planning workflow (Becker et al. 2009; Strodl 2007). The PLANETS preservation planning workflow specify the use of workshop style to elicit from stakeholders preservation requirements of any identified object type in a specific domain (Strodl 2007, sec. 4.2). This "requirements definition" is supposed to take into consideration institutional polices and stakeholders' goals of a preservation solution before listing these requirements down as high level goals tailing off into detail quantifiable objectives in a tree-like structure. Specifying the requirements of different stakeholders can indeed help determine which characteristics to preserve. However, such requirement identification though important, is not an objective-enough solution primarily because it is still difficult to translate these requirements into measurable terms. Becker et al. (2009) elaborate that translating "requirements definition" into preservation recommendations on what to preserve and determining the weightage of these requirements is still largely subjective in nature. The authors use the example of how curators' reluctance to quantify on what losses can be tolerated for example, would lead them to have the tendency to want to preserve everything (Becker et al. 2009, 142). The authors concluded that some resolution is still needed to make decision on what and how much to preserve more objective, automatic and measurable (Becker et al. 2009).

To make matters worst, there is contradiction in the literature on whether significant characteristics should be discretely binary i.e., should a significant characteristic be distinctly labeled as either to preserve or not to preserve, with no middle ground; or can a characteristic's significance be determined on a continuum and given a weightage? Dappert and Farquhar (2009, 303) assert that "significance is not absolute and binary," while Becker et al. (2009, 145) speak of the opposite, stating that "significant properties of digital objects are frequently seen as binary criteria that are either preserved or not."

This opens questions for further research. So, how then can we have an objective way to decide what characteristics to preserve?

# 7.    Recommendation: Make explicit at least some or one *functions/ purposes* of the object

Rothenberg (2000, 57) expresses that those properties to preserve "are derived entirely from the purpose that the entity is to serve." I first heard about the using an object's function in determining what to preserve from Martin Doerr. In his opinion, we are in a conundrum because of a fundamental flaw in how digital preservation is approached—the way discussions go seems to predicate that we have to predict (all) purposes the materials might serve in the future. In any case, the solutions we see so far—with the elaborate involvement of many stakeholders and specification of many requirements—seem to infer that. That is too huge a responsibility to assume. We can neither foresee accurately nor comprehensively the use or all uses of any materials in future. It is no wonder we are trapped and not able to identify any objective solutions. For practicality and the fact that resources are finite,[10] we therefore cannot be preserving blindly for all unknown amorphous purposes. The key is thus to *make explicit* at least one *purpose or function* of the object.

Though the idea of starting with the uses and functions of the object is not new—the most prominent being Rothenberg and Bikson's strategy [seen in the article by Knight and Pennock (2009)]—the need to articulate and make explicit at least one function of the object definitely is. Most solutions reviewed so far did not seem to place enough emphasis on the function (or intended use) of the object.

As a starting point on how to do that, the digital preservation community could perhaps take a leaf out of the archival world, more specifically, by employing what has already been so familiar to the archival world—the functional analysis (M. Doerr, pers. comm.). The International Council on Archives (ICA) has clear guidelines (as seen in *International Standard for Describing Functions*, 2007) on how one could articulate the functions which the records were designed to fulfill. This could then successfully put the record in the context of its creation, use and preservation.

The proposed gap-filler can be represented schematically in figure 1. This is just a formative recommendation and a lot more deliberation will be required.

---

[10] Many have corroborated on this point and argue the need to discriminate significant properties for practicality reasons. This include Hedstrom and Lee (2002), Ross and Hedstrom (2005), Rusbridge (2006), and more recently, Dappert and Farquhar (2009).

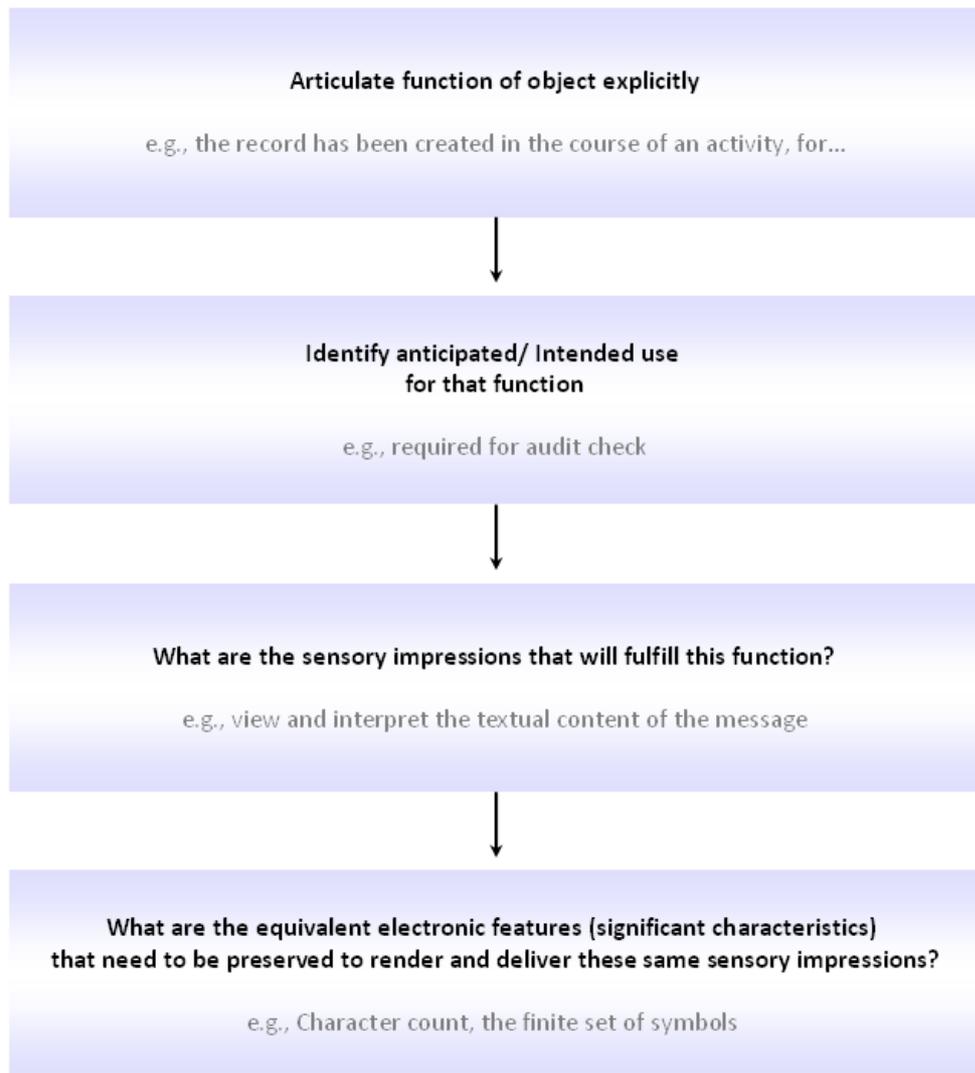

*Figure 1.* This figure illustrates steps in defining significant properties of digital objects starting with an object's function.

## 8.    Conclusion

In an attempt to answer the question of *what exactly should we preserve*, we have looked at how the digital preservation community and scholars address this question and how they have expounded on the concept of significant properties.

This review reveals that significant properties are properties that are both *inherent* as well as *given/ assigned* to a digital object. Thus, while there are tools to identify the inherent characteristics needed to render the object at some point in future (i.e. make it accessible), there is no easy way to identify the appraised significant properties because these properties are not inherent; they are waiting to be assigned! There is obviously a lack of objectivity to help stakeholders identify and decide what these appraised significant properties should be. Appraising significant properties of digital objects is still very much subjective. While "appraisal" (assessing and selecting what documents—including digital documents—are significant and worthwhile to preserve) is very much practice in the archival community, archival appraisal generally addresses what electronic records/ documents to preserve, but not what *features* on these documents to preserve. In other words, there is currently no formal way for the stakeholders to objectively appraise and thus give/assign significance that inform on what features to preserve. Until there is an objective solution to assign significance to specific features of an object rather than to the object in its current material or digital form, identification of a complete set of significant properties of any object will remain a difficult task.

One formative solution to that problem proposed by Martin Doerr is to articulate the *functions* of the digital object formally, in some *finite* way (i.e. explicate at least one function/ purpose of that object). Though functions and an object's possible uses are open-ended and infinite, our resources are not.

This thus raises another question for further contemplation: Whether it is too prudent for us to only identify one purpose/ function that the preservation action should cater for. Yeo (2010, 102) seems to imply that the preservation community in all eagerness to find a definitive solution should not ignore "the multiplicity of present and future users." He eloquently states that "affordances of records are wide-ranging . . . and affinitive" (Yeo 2010, 104). That means an object can have wide-ranging use to many communities, not just for the intended use that was the impetus to its creation. However, in the name of finite resources, even a current and active digital object is never everything to everyone. So why should a preserved digital object have to outdo its predecessors and coddle the unattainable by being everything to everyone? Such idealism, even if achievable would mean formidable investment and resources that will meet yet another challenge—sustainability. And since an object's multifaceted functions are—in the words of Yeo (2010)—affinitive, it will be reasonable and plausible to just earnestly identify a finite number of functions of the object for long-term preservation.